\documentclass[10pt,conference]{IEEEtran}
\usepackage[dvips]{graphicx}
\usepackage{indentfirst}
\usepackage{amsmath}
\usepackage{amssymb}
\usepackage{threeparttable}
\usepackage{multirow}
\usepackage{subfigure}
\usepackage{xcolor}
\usepackage{cite}
\usepackage{enumerate}
\usepackage{algorithm}
\usepackage{algorithmic}
\usepackage{setspace}
\usepackage{multicol}
\usepackage{stfloats}
\usepackage{array}

\newtheorem{Theorem}{Theorem}

\ifCLASSOPTIONcompsoc
\else
\fi

\ifCLASSINFOpdf
\else
\fi

\allowdisplaybreaks

\IEEEoverridecommandlockouts\IEEEpubid{\makebox[\columnwidth]{ 978-1-6654-3540-6/22/\$31.00 ~\copyright~2022 IEEE \hfill} \hspace{\columnsep}\makebox[\columnwidth]{ }}

\begin{document}

\title{Dynamic Power and Rate Allocation for NOMA Based Vehicle-to-Infrastructure Communications}

\author{Chongtao Guo and Bin Liao \\
College of Electronics and Information Engineering, Shenzhen University, Shenzhen 518060, Guangdong, China \\
    Email: ctguo@szu.edu.cn; binliao@szu.edu.cn
}

\maketitle

\begin{abstract}
In this paper, a non-orthogonal multiple access (NOMA) based downlink vehicle-to-infrastructure  network is considered. Particularly, we focus on the specific case of two users, one of which requires reliable road-safety-critical data transmission while the other pursues high-capacity services, with extension to multi-user scenarios. Leveraging only slow fading of channel state information, the transmit powers and target rates are jointly optimized to maximize the expected sum throughput of the capacity hungry user, with consideration of the payload delivery outage probability of the reliability sensitive user.
The optimization is formulated as an unconstrained single-objective sequential decision problem via introducing a dual variable.
A dynamic programming based algorithm is then designed to derive the optimal policy that maximizes the Lagrangian.
Afterwards, a bisection search based method is proposed to find the optimal dual variable. The proposed scheme is shown by numerical results to be superior to the baseline methods in terms of the expected return, performance region, and objective value.
\end{abstract}


\section{ Introduction }\label{Section_Introduction}

As a key enabler to intelligent transportation systems, vehicular communications take charge of information exchange among various entities on or near roads, including vehicle-to-infrastructure (V2I), vehicle-to-vehicle (V2V), vehicle-to-pedestrian (V2P), and vehicle-to-network (V2N) communications.
Depending upon the applications of the carried data, connections may be concerned with different quality of service (QoS), e.g., road-safety-critical data transmission usually requires high reliability and low latency while infortainment data traffic often desires high capacity \cite{2021-CST-ChallenSoluCellularV2X}.

To embrace the challenge of fast channel variation in high-mobility vehicular environment, slowly varying large-scale channel fading information has been utilized to develop smart wireless resource allocation to satisfy diverse QoS requirements of various links \cite{2016-TWC-ClusterBasedRAManageForD2DV2X, 2017-TCOM-ResourceAllocationForD2DV2X, 2019-TWC-RAforVehLowLatHighReliability, 2021-WCL-RAandBlockSelection}.
In \cite{2016-TWC-ClusterBasedRAManageForD2DV2X}, the transmit power and resource block are jointly allocated to maximize the throughput of cellular users with restrictions on the rate outage probability of V2V users.
For spectrum sharing between V2I and V2V links, the 
scheme in \cite{2017-TCOM-ResourceAllocationForD2DV2X} optimizes the sum capacity and minimum capacity of V2I links while ensuring the signal-to-interference-plus-noise ratio (SINR) outage probability of V2V connections.
The resource allocation in \cite{2019-TWC-RAforVehLowLatHighReliability} guarantees the average queueing latency and packet dropping probability of V2V links.
The work in \cite{2021-WCL-RAandBlockSelection} optimizes  channel, power, and blocklength allocation in V2X networks consisting of finite-blocklength V2V and infinite-blocklength V2N communications,
by minimizing the maximum latency of V2V links subject to constraints on the V2V links' rate outage probability and V2N links' ergodic capacity.

While resource allocation decision is made only once for each realization of large-scale fading in \cite{2016-TWC-ClusterBasedRAManageForD2DV2X, 2017-TCOM-ResourceAllocationForD2DV2X, 2019-TWC-RAforVehLowLatHighReliability, 2021-WCL-RAandBlockSelection}, resource management has been further designed to adapt to the variation of fast fading in \cite{2019-TVT-DRLbasedRAforV2V, 2019-JSAC-SpecSharingMultiAgent, 2021-TVT-MetaRLV2X, 2021-TVT-MRLenablesSpeAccess}.
The deep reinforcement learning based decentralized power level and subchannel allocation scheme proposed in \cite{2019-TVT-DRLbasedRAforV2V} provides lower latency for V2V links and higher capacity for V2I links.
Leveraging multi-agent reinforcement learning, the spectrum sharing and power allocation are designed to maximize the capacity of V2I links and improve data transmission  reliability of V2V links in \cite{2019-JSAC-SpecSharingMultiAgent}.
To promote the adaptability to fast environment variation,  a meta-reinforcement learning based resource allocation scheme is presented in \cite{2021-TVT-MetaRLV2X} to enhance the QoS of V2I and V2V links.
In \cite{2021-TVT-MRLenablesSpeAccess}, the sum throughput of V2I links is maximized with constraints on the latency and reliability of V2V links by multi-agent reinforcement learning based spectrum sharing.
Besides, the spectrum-efficient non-orthogonal multiple access (NOMA) technique has also been applied to V2X networks. For instance, a two-stage scheme of centralized spectrum allocation and distributed power control is devised in \cite{2021-IoTJ-CenRADisPC} to maximize the sum capacity of the group-cast system.
The user scheduling and power allocation are proposed to minimize the sum delay of V2I and V2V users via reinforcement learning in \cite{2021-INFOCOM-RAforLowLatNOMA}.
The power of the infrastructure and backscatters in \cite{2021-TITS-EERA6G} can maximize the total energy efficiency while guaranteeing the minimum data rate for all links. However, little attention has been paid to the tail behavior management of the road-safety-critical performance for dynamic resource allocation in NOMA based vehicular communications, where the tail behavior, posing  a significant threat to road safety, can be measured by the performance outage probability.
In particular, 3GPP imposes a requirement of transmitting a given amount of data within a predetermined time duration with sufficient success probability \cite{2018-Proc-URLLCTail, 2017-3GPP-22261}.

The aforementioned research gap motivates us to investigate the dynamic power and rate allocation for a NOMA based downlink V2I network, in which one user requires reliable payload delivery and the other pursues high capacity. More specifically, we maximize the expected data throughput of the capacity hungry user subject to a  data delivery outage constraint of the reliability sensitive user.
Different from most of the available works that do not capture the optimality of the dynamic resource allocation policy, we provide theoretical performance guarantee for the proposed algorithm from the perspective of optimization theory.
The main contributions are threefold.
First, a finite Markov decision process (MDP) with appropriate reward design is developed such that the transmitter, acting as the agent, can optimize the Lagrangian of the original problem by equivalently maximizing its expected return.
Second, a dynamic programming based algorithm is devised to maximize the Lagrangian.
Third, the dual variable is fast optimized by bisection search.

\section{System Model and Problem Formulation}\label{Sec:SystemModel}

In this section, we will introduce the network scenario, signal transmission model, and problem formulation successively.

\subsection{Network Scenario}
Consider a downlink V2I network consisting of one infrastructure access point (AP) and two vehicular users. The first user (${\rm U}_1$), requires reliable road-safety-related data transmission while the second user (${\rm U}_2$) is throughput hungry but reliability and latency insensitive.
$N$ data packets are required to be delivered from the AP to ${\rm U}_1$ within $T$ slots with outage probability no higher than $\delta$. The length of each slot is $\tau$, which can be regarded as the channel coherence time on the order of, say,  hundreds of microseconds in vehicular environment.
In this regard, the small-scale fading of the downlink V2I channels remains constant within each slot but varies fast from one slot to another.
To avoid substantial signalling overhead caused by channel state information (CSI) feedback in every slot, we assume that the AP has the statistical information rather than the realization of the fast fading in each slot.
However, the large-scale fading is considered to be available at the AP since it varies on a slow scale and can be fed back 
less frequently.
The channel power gain of the $k$th ($k\in\{1,2\}$ hereafter) downlink V2I connection in the $t$th slot is modeled as
\begin{equation}
h_k[t] = \beta_k g_k[t],
\end{equation}
where $\beta_k$ and $g_k[t]$ account for the large-scale fading and small-scale fading, respectively.
We consider Rayleigh fast fading in this article, i.e., $g_k[t]$ is independent and identically distributed exponential random variables with unit mean for all $k$ and $t$.

\subsection{Signal Transmission Mechanism}
Assume that the AP transmits signals to the two users over the same spectrum of bandwidth $W$ using the NOMA technique.
The superposition signal at the $t$th slot is constructed as
\begin{equation}
x[t] =  \sqrt{P V_1[t]} \cdot x_1[t] + \sqrt{P V_2[t]} \cdot x_2[t]
\end{equation}
where $P$ is the total power consumption, $x_k[t]$ is the normalized baseband signal of ${\rm U}_k$ with unit power, and $V_k[t]$ denotes the proportion of power allocated to ${\rm U}_k$ in the $t$th slot.
We consider $L$ discrete possible power allocation choices, forming the power set $\mathcal{V} = \{(V_1^l, V_2^l)|l=1,2,\cdots, L\}$, where each power allocation can make full utilization of the total power, i.e., $V_1^l + V_2^l =1$ for $l=1,2,\cdots, L$.
The received signal of ${\rm U}_k$ at the $t$th slot is
\begin{equation}
y_k[t] = h_k[t]x[t] + z_k[t] = \beta_k g_k[t]x[t] + z_k[t],
\end{equation}
where $z_k[t]$ with variance $\mathbb{E}[|z_k[t]|^2]=\sigma_k^2$ represents the additive white Gaussian noise (AWGN) and cochannel interference.

In downlink NOMA transmission with successive interference cancellation, the design of target transmission rates of ${\rm U}_1$ and ${\rm U}_2$, denoted by $R_1[t]$ and $R_2[t]$, respectively, is closely related to  the decoding order, $O[t]\in \mathcal{O} = \{ O_{1\rightarrow 2}, O_{2\rightarrow 1}\}$, where $O_{1\rightarrow 2}$ represents the order of decoding ${\rm U}_1$ and ${\rm U}_2$ successively and $O_{2\rightarrow 1}$ is similarly defined.
The target rates of ${\rm U}_k$, in the unit of the number of packets per slot, are selected from the discrete set $\mathcal{R}_k$, 
and the number of bits contained in each packet of ${\rm U}_k$ is denoted by $Y_k$.
In the following, we discuss the data transmission processes for the two decoding orders in the $t$th slot, where the power allocation is $(V_1[t], V_2[t])$, and the transmission rates of ${\rm U}_1$ and ${\rm U}_2$ are $R_1[t]$ and $R_2[t]$, respectively.

\subsubsection{$O[t]\! =\! O_{1\rightarrow 2}$}For this decoding order, ${\rm U}_1$ decodes its signal by treating the signal from ${\rm U}_2$ as interference. This leads to the following channel capacity:
\begin{equation}
R_1^{1\rightarrow 2, 1}[t] = \tau W Y_1^{-1} \log_2 \left( 1 + \frac{ P \beta_1  g_1[t] V_1[t] }{\sigma_1^2 +  P \beta_1 g_1[t] V_2 [t] }\right),
\end{equation}
which represents the number of packets that can be successfully carried in the $t$th slot.
Since the target rate $R_1[t]$ can be supported only if the channel capacity covers $R_1[t]$, the number of packets successfully received by ${\rm U}_1$ in the $t$th slot is thus given by
\begin{equation}\label{eqn:order1:D1t}
D_1[t] \! = \!
\begin{cases}
R_1[t],& {\rm if} \ R_1^{1\rightarrow 2, 1}[t] \ge \!  R_1[t]\\
0,     & {\rm otherwise}.
\end{cases}
\end{equation}

At ${\rm U}_2$, it first decodes the signal of ${\rm U}_1$ by taking its own signal as interference, with channel capacity
\begin{equation}
R_1^{1\rightarrow 2, 2}[t] = \tau W Y_1^{-1} \log_2 \left( 1 + \frac{ P \beta_2  g_2[t] V_1[t] }{\sigma_2^2 +  P \beta_2 g_2[t] V_2 [t] }\right).
\end{equation}
If $R_1[t]$ is covered by this achievable rate, ${\rm U}_2$ can decode its own signal suffering no interference from ${\rm U}_1$ with capacity
\begin{equation}\label{eqn:U2Capacity:case1}
R_2^{1\rightarrow 2, a}[t] = \tau W Y_2^{-1} \log_2 \left( 1 + \frac{ P \beta_2  g_2[t] V_2[t] }{\sigma_2^2  }\right).
\end{equation}
Otherwise, ${\rm U}_2$ has to decode its information by treating the signal from ${\rm U}_1$ as interference, giving rise to channel capacity
\begin{equation}\label{eqn:U2Capacity:case2}
R_2^{1\rightarrow 2, b}[t] = \tau W Y_2^{-1} \log_2 \left( 1 + \frac{ P \beta_2  g_2[t] V_2[t] }{\sigma_2^2 + P \beta_2 g_2[t] V_1 [t] }\right).
\end{equation}
For each case, the data of ${\rm U}_2$ can be successfully received at ${\rm U}_2$ only if the capacity in \eqref{eqn:U2Capacity:case1} or \eqref{eqn:U2Capacity:case2} is no less than $R_2[t]$.
Overall, the number of packets that can be successfully received by ${\rm U}_2$ in the $t$th slot, denoted by $D_2[t]$, can be figured out as
\begin{equation}\label{eqn:order1:D2t}
D_2[t] =
\begin{cases}
R_2[t], & {\rm if} \ R_1^{1\rightarrow 2, 2}[t] \ge  R_1[t], R_2^{1\rightarrow 2, a}[t] \ge  R_2[t]\\
R_2[t], & {\rm if} \ R_1^{1\rightarrow 2, 2}[t] <  R_1[t], R_2^{1\rightarrow 2, b}[t] \ge  R_2[t] \\
0,      &  {\rm otherwise}.
\end{cases}
\end{equation}

\subsubsection{ $O[t] = O_{2\rightarrow 1}$}Following the above analysis, we can derive the expressions of $D_2[t]$ and $D_1[t]$ for decoding order $O_{2\rightarrow 1}$ as

\vspace{-1em}
\begin{small}
\begin{equation}
D_2[t] \!=\!
\begin{cases}
R_2[t],&\!\! {\rm if} \ \tau  W Y_2^{-1} \log_2 \left( 1 + \frac{ P \beta_2  g_2[t] V_2[t] }{\sigma_2^2 \!+\!  P \beta_2 g_2[t] V_1 [t] }\right) \!\ge\!  R_2[t]\\
0,     &\!\! {\rm otherwise}
\end{cases}
\end{equation}
\end{small}
\vspace{-1em}

\noindent and

\vspace{-1em}
\begin{small}
\begin{equation}
D_1[t] \!=\!\!
\begin{cases}
\!R_1[t], &\!\! {\rm if} \ \tau W Y_2^{-1} \log_2 \left( 1 \!+\! \frac{ P \beta_1  g_1[t] V_2[t] }{\sigma_1^2 +  P \beta_1 g_1[t] V_1 [t] }\right) \!\ge\!  R_2[t], \\
        &\!\! \tau W Y_1^{-1} \log_2 \left( 1 \!+\! \frac{ P \beta_1  g_1[t] V_1[t] }{\sigma_1^2  }\right) \!\ge\!  R_1[t] \\
\!R_1[t], &\!\! {\rm if} \ \tau W Y_2^{-1} \log_2 \left( 1 \!+\! \frac{ P \beta_1  g_1[t] V_2[t] }{\sigma_1^2 +  P \beta_1 g_1[t] V_1 [t] }\right) \!<\!  R_2[t], \\
        &\!\! \tau W Y_1^{-1} \log_2 \left( 1 \!+\! \frac{ P \beta_1  g_1[t] V_1[t] }{\sigma_1^2 + P \beta_1 g_1[t] V_2 [t] }\right) \!\ge\!  R_1[t] \\
\!0,      &\!\!  {\rm otherwise},
\end{cases}
\end{equation}
\end{small}
respectively.

\subsection{Problem Statement}
It is seen from the above discussion that the data transmission processes of the two users are influenced by the transmit powers and target rates, which are closely related to the decoding order.
Therefore, the power, rate, and decoding order should be jointly determined for every possible system situation, also called state.

\subsubsection{State}
A state, $s$, can be characterized by $s = (E^s, Z^s)$ with $E^s$ and $Z^s$ being the numbers of remaining slots and remaining packets of ${\rm U}_1$, respectively.
Clearly, there are $T(N\!+\!1)+1$ possible states, forming the state space $\mathcal{S}$, among which the initial state is always $s_0=(T, N)$.
The states with nonzero remaining slots are  referred to as
nonterminal states, forming the set $\mathcal{S}^{-}=\{s| s \in \mathcal{S}, E^s > 0\}$ with cardinality $(T-1)(N+1)+1$.
Based on such definition, we can express the set of terminal states as $\mathcal{S}-\mathcal{S}^{-}$, with cardinality $N+1$.

\subsubsection{Action}
An action, $a$, can be defined as $ (O^a, V^a, R_1^a, R_2^a)$, where $O^a\in \mathcal{O}$,  $V^a\in \mathcal{V}$, $R_1^a\in \mathcal{R}_1$, and $R_2^a\in \mathcal{R}_2$ represent the decoding order, power, target rate of ${\rm U}_1$, and target rate of ${\rm U}_2$, respectively.
The set of actions, $\mathcal{A}$, can thus be expressed in the form of Cartesian product, i.e., $\mathcal{A} = \mathcal{O} \times \mathcal{V}  \times  \mathcal{R}_1 \times  \mathcal{R}_2$.

\subsubsection{Problem Formulation}

In this paper, we model the power, rate, and decoding order allocation as a policy optimization issue. Specifically, a policy $\pi(s)$ refers to a mapping from the set $\mathcal{S}^{-}$ to the set $\mathcal{A}$.
The policy optimization of maximizing the expected capacity of ${\rm U}_2$ subject to a constraint on the payload delivery outage probability of ${\rm U}_1$ is formulated as
\begin{equation} \label{OriProblem}
   \underset{  \pi \in \Pi  }{\max}      ~  {\mathbb{E}}_{\pi} \left[  \sum_{t=1}^T  D_2[t]  \right] \quad
{\rm s.t.}~ {\mathbb{P}}_{\pi} \left\{ \sum_{t=1}^T D_1[t] < N \right\} \le \delta,
\end{equation}
where $\Pi$ is the policy space with cardinality $|\Pi| = |\mathcal{A}|^{|\mathcal{S}^{-}|}=|\mathcal{A}|^{(T-1)(N+1)+1}$.
Since the number of possible policies grows exponentially with $T$ and $N$, exhaustive search is unserviceable in most practical systems.

\section{Proposed Approach}

In this section, we first derive the Lagrangian of the problem in \eqref{OriProblem} by introducing a Lagrange dual variable.
Then, we construct a finite MDP and figure out the optimal policy to maximize the Lagrangian by dynamic programming.
Finally, a bisection search based method is proposed to obtain the optimal dual variable.

\subsection{Solution Structure}

Associating a non-negative dual variable $\lambda$ with the constraint yields the following Lagrangian of the primal problem:
\begin{equation}\label{eqn:Lagrangian}
L(\pi, \lambda) \!=\! {\mathbb{E}}_{\pi} \!\left[  \sum_{t=1}^T  D_2[t]  \right] \!+ \lambda \left[\delta \!-\!  {\mathbb{P}}_{\pi}\! \left\{ \sum_{t=1}^T D_1[t] < N \right\}  \right].
\end{equation}
The Lagrange dual function is thus given by
\begin{equation}
f(\lambda)  = \underset{\pi}{\max} \  L(\pi, \lambda),
\end{equation}
with $\pi_{\lambda} = \arg\underset{\pi}{\max} \ L(\pi, \lambda)$ being the policy that maximizes the Lagrangian under fixed dual variable $\lambda$.
Afterwards, we find the best dual variable, $\lambda^*$, by handling the Lagrange dual problem
\begin{equation} \label{DualProblem}
   \underset{  \lambda  }{\min}     \ f(\lambda) \quad \quad \quad
{\rm s.t.}                        \ \lambda \ge 0.
\end{equation}
Finally, the policy $\pi_{\lambda^*}$ will be returned as the solution.

\subsection{Policy Optimization via Dynamic Programming}

\setcounter{equation}{25}
\begin{table*}
\begin{small}
\begin{equation}\label{eq:order1:D2dist}
\begin{split}
   \ \mathbb{P} \left\{ D_2[t] \!=\! R_2[t] \right\}
= \!\! \begin{cases}
{\rm e}^{-\varphi_3},   & {\rm if} \ V_1[t] - V_2[t] \left(2^{\frac{Y_1R_1[t]}{\tau W} } -1\right) > 0, V_2[t] - V_1 [t] \left( 2^{\frac{Y_2 R_2[t]}{\tau W}} - 1 \right) > 0, \varphi_2 > \varphi_1  \\
 {\rm e}^{-\varphi_2} + {\rm e}^{-\varphi_3}- {\rm e}^{-\varphi_1}, & {\rm if} \ V_1[t] - V_2[t] \left(2^{\frac{Y_1R_1[t]}{\tau W} } -1\right) > 0, V_2[t] - V_1 [t] \left( 2^{\frac{Y_2 R_2[t]}{\tau W}} - 1 \right) > 0, \varphi_2 \le \varphi_1 \\
{\rm e}^{-\varphi_3},   & {\rm if} \ V_1[t] - V_2[t] \left(2^{\frac{Y_1R_1[t]}{\tau W} } -1\right) > 0, V_2[t] - V_1 [t] \left( 2^{\frac{Y_2 R_2[t]}{\tau W}} - 1 \right) \le 0   \\
{\rm e}^{-\varphi_2}, & {\rm if} \ V_1[t] - V_2[t] \left(2^{\frac{Y_1R_1[t]}{\tau W} } -1\right) \le 0, V_2[t] - V_1 [t] \left( 2^{\frac{Y_2 R_2[t]}{\tau W}} - 1 \right) > 0 \\
0,  & {\rm if} \ V_1[t] - V_2[t] \left(2^{\frac{Y_1R_1[t]}{\tau W} } -1\right) \le 0, V_2[t] - V_1 [t] \left( 2^{\frac{Y_2 R_2[t]}{\tau W}} - 1 \right) \le 0
\end{cases}
\end{split}
\end{equation}
\end{small}
\rule{\textwidth}{0.2mm}
\end{table*}

As the optimal solution of an unconstrained sequential decision problem for a given $\lambda$, the policy $\pi_{\lambda}$ can be derived by constructing and solving  a finite MDP, where an agent interacts with its environment during the MDP.
Specifically, the agent, i.e., the AP, observes a state, $S_t$, at time $t$ from the state space, $\mathcal{S}^{-}$, and on that basis chooses an action, $A_{t+1}$, from the action space, $\mathcal{A}$, based on a policy, $\pi$.
One time slot later, the environment responds to the action taken in the previous slot by presenting a new state, $S_{t+1}$, and offering a reward, $R_{t+1}$, to the agent.
It is worth noting that $S_t$ denotes the state at the end of the $t$th slot, $R_t$ is the reward received at the end of the $t$th slot, and $A_t$ represents the action taken in the $t$th slot, for $t=\{1,2,3,\cdots, T\}$.
By defining $S_0=s_0$ as the initial state observed at the beginning of the first slot, we have the episodic trajectory of the agent-environment interaction given as $S_0, A_1, R_1, S_1, A_2, R_2, S_2, \cdots, A_T, R_T, S_T$. The reward $R_t$ is designed as 
\setcounter{equation}{15}
\begin{equation}\label{eqn:rewardDesign}
R_t =  D_2[t] +  \lambda c_t,
\end{equation}
where $c_t$ is given by
\begin{equation}
c_t =
\begin{cases}
0,               & {\rm if}   \ S_{t} \in \mathcal{S}^{-} \\
\delta -1,       & {\rm if}   \ S_{t} \in \mathcal{S} - \mathcal{S}^{-}, \ Z^{S_{t}} > 0 \\
\delta,          & {\rm if}   \ S_{t} \in \mathcal{S} - \mathcal{S}^{-}, \  Z^{S_{t}} = 0.
\end{cases}
\end{equation}
Since $R_t$ and $S_t$ have discrete probability distribution dependent only on the preceding state and action, the dynamics of the finite MDP can be characterized by
\begin{equation}
p(s', r| s, a) =  {\mathbb{P}} \left\{ S_t = s', R_t = r | S_{t-1} = s, A_t = a \right\},
\end{equation}
which denotes the probability of state $s'$ and reward $r$ at time $t$ given the preceding state $s$ at time $t-1$ and action $a$ at time $t$, where $s'\in\mathcal{S}$, $r\in\mathcal{R}$, $s\in\mathcal{S}^{-}$, and $a\in\mathcal{A}$.

The value of state $s$ under policy $\pi$ represents the expected return starting from $s$ and following $\pi$ thereafter, i.e.,
\begin{equation}
v_{\pi}(s) = {\mathbb{E}}_{\pi} \left[ \sum\nolimits_{j=1}^{E^s} R_{T - E^s + j} \middle\vert  S_{T-E^s} =s \right]
\end{equation}
for all $s\in\mathcal{S}^{-}$  and $v_{\pi}(s) = 0$ for all $s\in\mathcal{S} -\mathcal{S}^{-}$.
Since the reward design  is related to $\lambda$, the optimal policy  is denoted by $\pi^{\dag}(\lambda)$, which leads to the highest state values for all states, i.e., $v_{\pi^{\dag}(\lambda)}(s) \ge v_{\pi'}(s)$ holds for all $s\in\mathcal{S}$ and $\pi'\in\Pi$.
The following theorem provides an appealing property of $\pi^{\dag}(\lambda)$.
\begin{Theorem}\label{Theorem}
$\pi^{\dag}(\lambda)=\pi_{\lambda}$ if the MDP has fixed dynamics.
\end{Theorem}
\begin{IEEEproof}
We deploy policy $\pi$ for $M$ independent episodes with fixed dynamics, i.e., $p(s', r| s, a)$ keeps fixed for all $s'\in\mathcal{S}$, $r\in\mathcal{R}$, $s\in\mathcal{S}^{-}$, and $a\in\mathcal{A}$.
Then, the expected return of the initial state, $s_0$, can be derived by averaging the return realizations of the $M$ episodes as $M$ approaches infinity.
Let $I_{m}$ be the payload delivery success indicator of ${\rm U}_1$ in the $m$th episode under policy $\pi$, such that $I_{m}=1$ if the transmission is successful and $I_{m}=0$ otherwise.
Further, the reward received at time $t$ during the $m$th episode is notated as $R_{t, m}$, and $D_{k,m}(t)$ denotes the number of packets transmitted in the $t$th slot for ${\rm U}_k$ during the $m$th episode under policy $\pi$.
Finally, the state value of $s_0$ is figured out as

\begin{footnotesize}
\begin{equation}
\begin{split}
  &\ v_{\pi}(s_0)= \mathbb{E}_{\pi} \left[ \sum_{t=1}^T R_t\right] = \lim_{M\rightarrow\infty} \frac{1}{M} \sum_{m=1}^{M} \sum_{t=1}^{T} R_{t, m} \\
&=  \ \lim_{M\rightarrow\infty} \left[\frac{1}{M} \sum_{m=1}^{M} \sum_{t=1}^{T} D_{2, m}(t) \right] \\
  & \ \ ~~+  \lambda \cdot \lim_{M\rightarrow\infty} \left[\frac{1}{M} \sum_{m=1}^{M} \left[\delta I_m + (\delta-1)(1-I_m)\right] \right] \\
&=  \ \lim_{M\rightarrow\infty} \left\{\! \left[\frac{1}{M} \!\!\sum_{m=1}^{M} \sum_{t=1}^{T} D_{2, m}(t) \right] \!+\!  \lambda \left[ \frac{\sum_{m=1}^M I_m}{M} - (1\!-\!\delta) \right] \!\right\} \\
&=  \ {\mathbb{E}}_{\pi} \left[  \sum_{t=1}^T  D_2[t]  \right] + \lambda \left[\delta - {\mathbb{P}}_{\pi} \left\{ \sum_{t=1}^T D_1[t] < N \right\}  \right] \\
&=  \ L(\pi, \lambda).
\end{split}
\end{equation}
\end{footnotesize}

Thus, the maximizer of the left-hand-side expression, $\pi^{\dag}(\lambda)$, is also the maximizer of the right-hand-side expression, $\pi_{\lambda}$.
\end{IEEEproof}

According to Theorem \ref{Theorem},  $\pi_{\lambda}$ can be derived by tackling the finite MDP.
Generally, as long as the dynamics, $p(s', r| s, a)$, are available, $\pi_{\lambda}$ can be computed by dynamic programming \cite{2018-Book-ReinforcementLearning} with value iteration shown in the \textbf{Algorithm}, where the state value $V(s)$ can be expressed in a tabular fasion.
Upon the agent taking action $a$ in the $t$th slot after observing state $s$, the distribution of the next state $s'$ and received reward $r$ can be directly obtained from the distribution of $D_1[t]$ and $D_2[t]$. Assuming the action taken in the $t$th slot is $(O_{1 \rightarrow 2}, (V_1[t], V_2[t]), R_1[t], R_2[t])$, we will discuss the distribution of $D_1[t]$ and $D_2[t]$ for a general case.

\begin{table}[!t]
\vspace{-0.5em}
\begin{algorithm}[H]
\caption{Dynamic Programming for Obtaining $\pi_{\lambda}$} \label{Algo:PolicyOptGivenLambda}
\begin{algorithmic}[1]
\begin{footnotesize}

\STATE \textbf{Initialization:}

\begin{itemize}
\item  Initialize the dual variable, $\lambda$, and the error tolerance, $\xi$
\item  Set $V(s)=0$ for all $s\in\mathcal{S}$
\end{itemize}

\REPEAT

\STATE  $\Delta = 0$

\FOR {$s\in\mathcal{S}^{-}$}

\STATE  $v=V(s)$

\STATE  $V(s) = \max\limits_{a\in\mathcal{A}}  \sum\limits_{s'\in\mathcal{S}} \sum\limits_{r\in\mathcal{R}} p(s',r|s,a)[r+V(s')]$

\STATE  $\Delta = \max\{\Delta, |V(s) - v|\}$

\ENDFOR

\UNTIL {$\Delta < \xi$}

\STATE $v_*(s)=V(s)$ for all $s\in\mathcal{S}$

\STATE \textbf{Return:}  policy $\pi_{\lambda}$ with $\pi_{\lambda}(s) \!=\! \arg \underset{a}{\max}\!\sum_{s'\in\mathcal{S}} \sum_{r\in\mathcal{R}} p(s',r|s,a) [r + v_{*}(s')]$

\end{footnotesize}
\end{algorithmic}
\end{algorithm}
\vspace{-2.5em}
\end{table}

According to \eqref{eqn:order1:D1t}, $D_1[t]$ can take two values, i.e., $R_1[t]$ and $0$.
Leveraging the exponential distribution of $g_1[t]$ with unit mean, we have
\begin{equation}\small
\begin{split}
  & \ \mathbb{P} \left\{ D_1[t] = R_1[t] \right\}
=   \mathbb{P} \left\{ \frac{ P \beta_1  g_1[t] V_1[t] }{\sigma_1^2 +  P \beta_1 g_1[t] V_2 [t] }  \ge   2^{\frac{Y_1 R_1[t]}{\tau W}} - 1 \right\} \\
&=  \begin{cases}
\exp(- \varphi_0),   & {\rm if} \ V_1[t] - V_2 [t] \left( 2^{\frac{Y_1 R_1[t]}{\tau W}} - 1 \right) > 0 \\
0,   & {\rm otherwise},
\end{cases}
\end{split}
\end{equation}
where $\varphi_0$  is given by
\begin{equation}
\varphi_0 = \frac{\sigma_1^2\left( 2^{\frac{Y_1 R_1[t]}{\tau W}} - 1 \right)}{P \beta_1 \left[ V_1[t] - V_2 [t] \left( 2^{\frac{Y_1 R_1[t]}{\tau W}} - 1 \right) \right]}.
\end{equation}
Similarly, according to \eqref{eqn:order1:D2t}, $D_2[t]$ is either $R_2[t]$ or $0$.
Exploiting the exponential distribution of $g_2[t]$ with unit mean and defining $\varphi_1$, $\varphi_2$, and $\varphi_3$ as
\begin{align}
\varphi_1 & =  \frac{\sigma_2^2 \left(2^{\frac{Y_1R_1[t]}{\tau W} } -1\right)}{P \beta_2  \left[ V_1[t] -  V_2[t] \left(2^{\frac{Y_1R_1[t]}{\tau W} } -1\right) \right]} \\
\varphi_2 & =  \frac{\sigma_2^2\left( 2^{\frac{Y_2 R_2[t]}{\tau W}} - 1 \right)}{P \beta_2  \left[ V_2[t] - V_1 [t] \left( 2^{\frac{Y_2 R_2[t]}{\tau W}} - 1 \right) \right]} \\
\varphi_3 & =  \max\left\{ \varphi_1, \frac{\sigma_2^2\left(2^{\frac{Y_2 R_2[t]}{\tau W}}-1 \right)}{P \beta_2 V_2[t]} \right\},
\end{align}
we can derive $\mathbb{P} \left\{ D_2[t] = R_2[t] \right\}$ shown in \eqref{eq:order1:D2dist} on the top of the previous page, where the detailed deduction is omitted here due to space limitation.
Then, we have ${\mathbb{P}} \left\{ D_1[t] = 0\right\} = 1 - {\mathbb{P}} \left\{ D_1[t] = R_1[t]\right\}$ and ${\mathbb{P}} \left\{ D_2[t] = 0\right\} = 1 - {\mathbb{P}} \left\{ D_2[t] = R_2[t]\right\}$.

From the above analysis, the dynamics, $p(s', r| s, a)$, can be derived for decoding order $O_{1 \rightarrow 2}$ in general cases.
For $O_{2 \rightarrow 1}$ and special cases such as $R_1[t]=0$ and/or $R_2[t]=0$, $p(s', r| s, a)$ can also be easily obtained using similar approaches.
It is then straightforward to compute the expected capacity of ${\rm U}_2$  and the payload delivery outage probability of ${\rm U}_1$. Details for this are not provided here due to space limitation.

\subsection{Dual Variable Optimization}

According to the optimization theory, the dual problem in \eqref{DualProblem} is always convex.
In addition, as $\lambda$ increases from zero to infinity, policy $\pi_{\lambda}$ achieves nonincreasing payload delivery outage probability of ${\rm U}_1$ and expected capacity of ${\rm U}_2$.
Thus, the dual problem is equivalent to the minimization of $\lambda$ while satisfying the reliability requirement of ${\rm U}_1$, which can be efficiently addressed by bisection search.
In particular, we initialize $\lambda_{\min}$ and $\lambda_{\max}$ as the lower and upper bounds of the search range, respectively, where zero is assigned to $\lambda_{\min}$ and a sufficiently large number is allocated to $\lambda_{\max}$ in general.
Then, we look into the policy $\pi_{\lambda_0}$ with $\lambda_0=(\lambda_{\min}+\lambda_{\max})/2$.
In particular, we set $\lambda_{\min}=\lambda_0$ if the outage probability of ${\rm U}_1$ is greater than $\delta$ under $\pi_{\lambda_0}$ and $\lambda_{\max}=\lambda_0$ otherwise.
After continuously checking the middle point and narrowing down the search range, we have $\lambda^* = \lambda_{\max}$ when $|\lambda_{\max}-\lambda_{\min}|< \epsilon$, where $\epsilon$ is the error tolerance. It can be observed that such procedure has a low complexity  ${\rm{O}}(\log(1/\epsilon))$.

\section{Extension to Multi-user Cases}

The two-user model addressed above moderates the complexity brought about by the user pairing issue, allows us to extract neat and handy analytical results, and serves as a manageable starting point for the problem investigated in this paper.
The proposed power and rate allocation framework can be directly extended to the multi-user situation with $N_{c}$ capacity hungry users and $N_{r}$ reliability sensitive users.
In particular, one can construct the corresponding optimization problem of maximizing sum capacity of the $N_{c}$ users with $N_{r}$ constraints, form the Lagrangian by introducing $N_{r}$ Lagrange multipliers, design reward function following from \eqref{eqn:rewardDesign}, and derive the best power and rate allocation as well as decoding order by the proposed algorithm. Then, the primal-dual based method can be applied to optimize the Lagrange multipliers.
However, too many users necessitate the optimization of user pairing and increase the number of possible decoding orders, which leads to exponentially increasing action space.
Towards this end, one may employ a deep neural network (DNN) to characterize the state value rather than the tabular one in the proposed algorithm for the two-user case.
Then, deep Q Learning can be utilized to find the optimal policy, where the stability and convergence of the DNN training deserves special attention.
Besides, the iteration of Lagrange multipliers in multi-user cases will become much more complicated than the two-user case, e.g., the step size and convergence criterion need to be carefully designed.
Finally, it is also straightforward to extend this model-based approach to the model-free data-driven situation, where the distribution of the small-scale fading is unknown.
The detailed extension will be elaborated in our future work.

\section{Numerical Results}

\begin{figure}
\centering
\includegraphics[height=5.8cm]{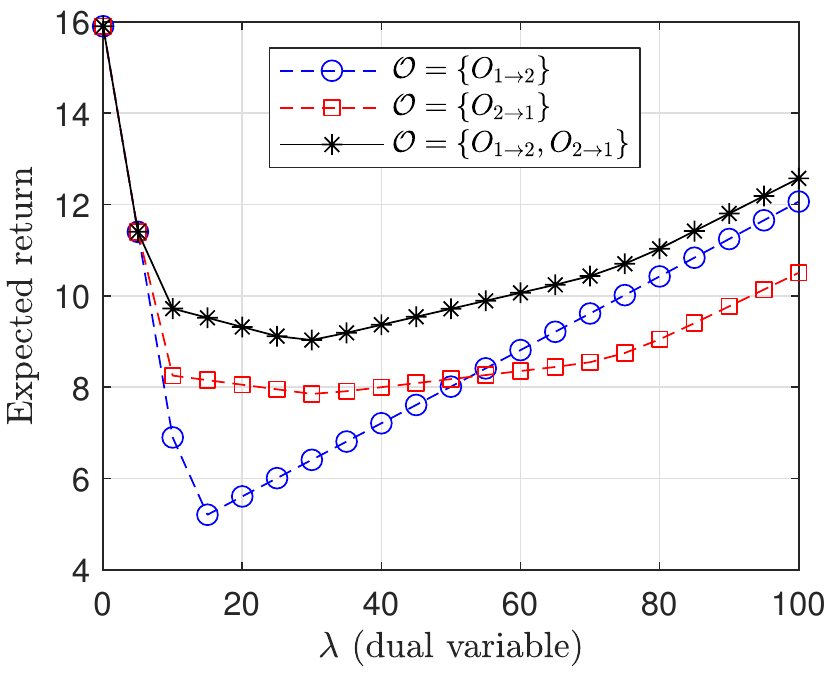}
\caption{Expected return of $\pi_{\lambda}$ with different dual variable $\lambda$. \label{Fig1}}
\end{figure}

Consider a specific experiment with $T=4$, $N=13$, $\tau=1$ ms, $W=1$ MHz, $P=30$ dBm, $\beta_1 = \beta_2=10^{-6}$, $Y_1 = Y_2 = 1,500$ bits, $\sigma_1^2 = \sigma_2^2 = -70$ dBm, and $\delta=0.1$.
The power allocation and target rate spaces are set as $\mathcal{V}=\{(0,0)$, $(0,1)$, $(0.1, 0.9)$, $(0.3, 0.7)$, $(0.5, 0.5)$, $(0.7, 0.3)$, $(0.9, 0.1)$, $(1, 0)\}$ and $\mathcal{R}_1 = \mathcal{R}_2 = \{0,1,2,3,4\}$.
Fig. \ref{Fig1} demonstrates the superiority of the proposed strategy from the perspective of the expected return achieved by $\pi_{\lambda}$.
In particular, the joint optimization of power, rate, and decoding order achieves much higher return than the approaches with a fixed decoding order.
Since the expected return of the optimal policy for any given $\lambda$ is the same as the $f(\lambda)$, it also validates the convexity of the dual function.

\begin{figure}
\centering
\includegraphics[height=5.8cm]{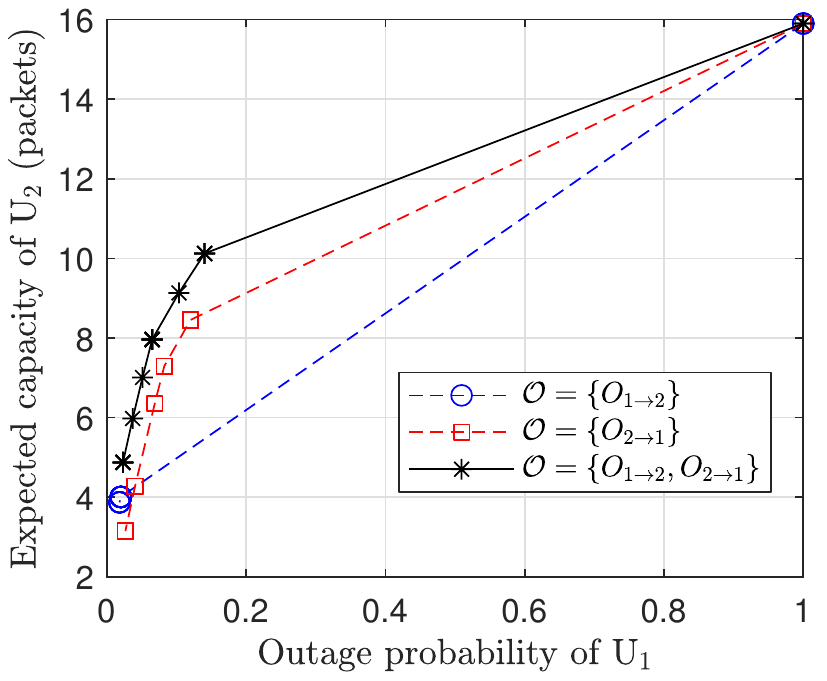}
\caption{Outage probability of ${\rm U}_1$ and expected capacity of ${\rm U}_2$  under $\pi_{\lambda}$ by varying the dual variable $\lambda$ from $0$ to $\infty$. \label{Fig2}}
\end{figure}

To further look into the achieved performance, Fig. \ref{Fig2} shows the performance region of the strategies with and without decoding order optimization.
As seen from Fig. \ref{Fig2}, the proposed scheme is more efficient in the sense that the interested performance metrics of the two users can be simultaneously enhanced when switching the policy from power and rate optimization to the proposed one with additional decoding order optimization.

\begin{figure}
\centering
\includegraphics[height=5.8cm]{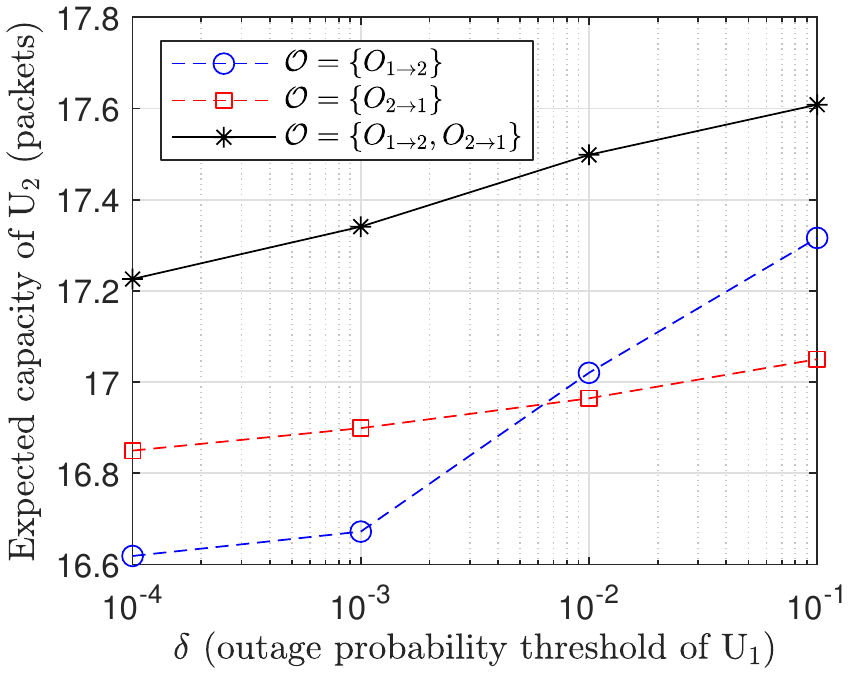}
\caption{Expected capacity of ${\rm U}_2$ with different thresholds of the payload delivery outage probability of ${\rm U}_1$. \label{Fig3}}
\end{figure}

Let $d_k$ denote the distances between the AP and ${\rm U}_k$, which are uniformly distributed in the range from $10$ m to $100$ m. The large-scale fading is modeled as $\beta_k = 10^{-3} d_k^{-2}$.  
By conducting 100 random realizations, Fig. \ref{Fig3} shows the average expected capacity of ${\rm U}_2$ with different reliability requirement of ${\rm U}_1$, where $T=10$, $N=16$, $Y_1 = Y_2 = 1,650$, $\mathcal{R}_1 = \mathcal{R}_2$ $= \{0,1,2\}$, and other parameters are set as before.
It can be observed that the proposed algorithm leads to higher capacity comparing with the schemes optimizing power and rate but not decoding order.
This is because the optimal decoding order may be different in different states and in different user distributions, making performance degradation under any fixed order.
Moreover, the capacity of ${\rm U}_2$ gets better as $\delta$ increases since the system could tilt in favor of ${\rm U}_2$ as the QoS requirement of ${\rm U}_1$ gets weaker.

\section{Conclusion}

A power and rate allocation algorithm has been developed for a NOMA based V2I network with diverse QoS requirements.
A finite MDP with appropriate reward design is constructed such that the agent can maximize the Lagrangian of the primal problem by optimizing its expected return in the agent-environment interaction.
Then, a low-complexity bisection search based method is proposed to solve the dual problem.
Finally, the superiority of the developed strategy to the baseline approaches has been validated and extensions of this work are also emphasized.

\section{Acknowledgment}

This work was supported in part by the Foundation of Shenzhen under Grant JCYJ20190808114213987, in part by the Department of Education of Guangdong Province under Grant 2018KTSCX195,  in part by Guangdong Basic and Applied Basic Research Foundation under Grant 2022A1515010188, and in part by the National Natural Science Foundation of China under Grant 62101340 and Grant 62171292.


\bibliographystyle{IEEEtran}
\bibliography{IEEEabrv,Reference_currentWork}


\end{document}